\documentclass[11pt, oneside]{article}   	
\usepackage{geometry}                	
\usepackage{pgfplots}
\pgfplotsset{compat=newest}
\usepgfplotslibrary{groupplots}
\usepgfplotslibrary{polar}
\usepgfplotslibrary{smithchart}
\usepgfplotslibrary{statistics}
\usepgfplotslibrary{dateplot}
\usetikzlibrary{arrows.meta}
\usetikzlibrary{backgrounds}
\usepgfplotslibrary{patchplots}
\usepgfplotslibrary{fillbetween}
\pgfplotsset{%
layers/standard/.define layer set={%
    background,axis background,axis grid,axis ticks,axis lines,axis tick labels,pre main,main,axis descriptions,axis foreground%
}{grid style= {/pgfplots/on layer=axis grid},%
    tick style= {/pgfplots/on layer=axis ticks},%
    axis line style= {/pgfplots/on layer=axis lines},%
    label style= {/pgfplots/on layer=axis descriptions},%
    legend style= {/pgfplots/on layer=axis descriptions},%
    title style= {/pgfplots/on layer=axis descriptions},%
    colorbar style= {/pgfplots/on layer=axis descriptions},%
    ticklabel style= {/pgfplots/on layer=axis tick labels},%
    axis background@ style={/pgfplots/on layer=axis background},%
    3d box foreground style={/pgfplots/on layer=axis foreground},%
    },
}

\pgfplotsset{
colormap={plots1}{rgb=(0.00146200,0.00046600,0.01386600)
rgb=(0.01656100,0.01313600,0.08028200)
rgb=(0.05164400,0.03247400,0.15925400)
rgb=(0.09299000,0.04558300,0.23435800)
rgb=(0.14907300,0.04546800,0.31708500)
rgb=(0.21109500,0.03703000,0.37856300)
rgb=(0.27134700,0.04092200,0.41197600)
rgb=(0.32892100,0.05782700,0.42751100)
rgb=(0.37900100,0.07625300,0.43271900)
rgb=(0.43498700,0.09706900,0.43203900)
rgb=(0.49102200,0.11717900,0.42555200)
rgb=(0.54715700,0.13692900,0.41351100)
rgb=(0.60313900,0.15715100,0.39589100)
rgb=(0.65236900,0.17642100,0.37558600)
rgb=(0.70650000,0.20072800,0.34777700)
rgb=(0.75842200,0.22909700,0.31526600)
rgb=(0.80708200,0.26269200,0.27889800)
rgb=(0.84670900,0.29755900,0.24411300)
rgb=(0.88630200,0.34258600,0.20296800)
rgb=(0.91987900,0.39338900,0.16007000)
rgb=(0.94696500,0.44919100,0.11527200)
rgb=(0.96732200,0.50907800,0.06865900)
rgb=(0.97966600,0.56505700,0.03140900)
rgb=(0.98696400,0.63048500,0.03090800)
rgb=(0.98712400,0.69794400,0.08773100)
rgb=(0.98003200,0.76683700,0.16635300)
rgb=(0.96624300,0.83619100,0.26153400)
rgb=(0.95154600,0.89622600,0.36562700)
rgb=(0.94954500,0.95506300,0.50786000)
rgb=(0.98836200,0.99836400,0.64492400)},
}
\pgfplotsset{
colormap={plots1}{rgb=(0.00146200,0.00046600,0.01386600)
rgb=(0.01656100,0.01313600,0.08028200)
rgb=(0.05164400,0.03247400,0.15925400)
rgb=(0.09299000,0.04558300,0.23435800)
rgb=(0.14907300,0.04546800,0.31708500)
rgb=(0.21109500,0.03703000,0.37856300)
rgb=(0.27134700,0.04092200,0.41197600)
rgb=(0.32892100,0.05782700,0.42751100)
rgb=(0.37900100,0.07625300,0.43271900)
rgb=(0.43498700,0.09706900,0.43203900)
rgb=(0.49102200,0.11717900,0.42555200)
rgb=(0.54715700,0.13692900,0.41351100)
rgb=(0.60313900,0.15715100,0.39589100)
rgb=(0.65236900,0.17642100,0.37558600)
rgb=(0.70650000,0.20072800,0.34777700)
rgb=(0.75842200,0.22909700,0.31526600)
rgb=(0.80708200,0.26269200,0.27889800)
rgb=(0.84670900,0.29755900,0.24411300)
rgb=(0.88630200,0.34258600,0.20296800)
rgb=(0.91987900,0.39338900,0.16007000)
rgb=(0.94696500,0.44919100,0.11527200)
rgb=(0.96732200,0.50907800,0.06865900)
rgb=(0.97966600,0.56505700,0.03140900)
rgb=(0.98696400,0.63048500,0.03090800)
rgb=(0.98712400,0.69794400,0.08773100)
rgb=(0.98003200,0.76683700,0.16635300)
rgb=(0.96624300,0.83619100,0.26153400)
rgb=(0.95154600,0.89622600,0.36562700)
rgb=(0.94954500,0.95506300,0.50786000)
rgb=(0.98836200,0.99836400,0.64492400)},
}
\pgfplotsset{
colormap={plots1}{rgb=(0.00146200,0.00046600,0.01386600)
rgb=(0.01656100,0.01313600,0.08028200)
rgb=(0.05164400,0.03247400,0.15925400)
rgb=(0.09299000,0.04558300,0.23435800)
rgb=(0.14907300,0.04546800,0.31708500)
rgb=(0.21109500,0.03703000,0.37856300)
rgb=(0.27134700,0.04092200,0.41197600)
rgb=(0.32892100,0.05782700,0.42751100)
rgb=(0.37900100,0.07625300,0.43271900)
rgb=(0.43498700,0.09706900,0.43203900)
rgb=(0.49102200,0.11717900,0.42555200)
rgb=(0.54715700,0.13692900,0.41351100)
rgb=(0.60313900,0.15715100,0.39589100)
rgb=(0.65236900,0.17642100,0.37558600)
rgb=(0.70650000,0.20072800,0.34777700)
rgb=(0.75842200,0.22909700,0.31526600)
rgb=(0.80708200,0.26269200,0.27889800)
rgb=(0.84670900,0.29755900,0.24411300)
rgb=(0.88630200,0.34258600,0.20296800)
rgb=(0.91987900,0.39338900,0.16007000)
rgb=(0.94696500,0.44919100,0.11527200)
rgb=(0.96732200,0.50907800,0.06865900)
rgb=(0.97966600,0.56505700,0.03140900)
rgb=(0.98696400,0.63048500,0.03090800)
rgb=(0.98712400,0.69794400,0.08773100)
rgb=(0.98003200,0.76683700,0.16635300)
rgb=(0.96624300,0.83619100,0.26153400)
rgb=(0.95154600,0.89622600,0.36562700)
rgb=(0.94954500,0.95506300,0.50786000)
rgb=(0.98836200,0.99836400,0.64492400)},
}

\usepackage{amsthm,hyperref,amsmath}
\newtheorem{definition*}{Defintion}

\title{Same-Score Streaks: A Case Study in Probability Modeling}
\author{Peter Staab and Rick Cleary}
\begin{document}
\maketitle

\section{A Problem}

Modeling the probability of rare events is a challenging, intriguing and sometimes economically or socially important problem in mathematics.  Rare events like earthquakes, severe storms or global pandemics have enormous consequences for individuals, organizations and governments.  Estimating the small but real probabilities of disasters is a staple of the science of risk assessment, a field where many college math majors find careers.  

Applications in risk assessment often require substantial subject area knowledge or collaboration with experts in the natural sciences, finance or public health. This makes it hard for students to have access to the area.  Fortunately, the principles of such modeling can be explored with more familiar, if less consequential, topics.  This paper suggest ways to approach one such question using probability modeling, historical data and simulations related to outcomes of basketball and baseball games.

In early 2013, co-author Rick Cleary received an email from a sports reporter at a small newspaper in North Carolina.  The reporter’s beat included covering the men’s basketball team at Barton College where the team had just won three consecutive games by identical scores.

\begin{center}
\begin{tabular}{l|ll}
January 21, 2013 & Barton 76 & Pfeiffer 68 \\
January 24, 2013 & Barton 76 &  Queens 68 \\
January 26, 2013 & Barton 76 & Erskine 68\\
\end{tabular}
\end{center}

The reported asked what were the chances of identical scores in three consecutive college basketball games?  Prof. Cleary received this email because he had some experience modeling rare events involving streaks in sports \cite{cleary:2010}.  His immediate answer was “I don’t know, but we can approximate it if we make some assumptions.”  

\section{A first crack at a solution }

At the time this question came to light, Prof. Cleary was teaching about 60 students in a course on Sports Applications of Mathematics, and in a happy coincidence they were covering a first unit on probability. So let’s pose to readers the question Prof. Cleary posed to his class:  How would you answer the reporter’s question?  (Take some time and think about it!  We’ll wait…)

When faced with a problem like this, the first step is to make some assumptions that let us get a handle on the problem.  This in itself is a mathematical balancing act:  Too many assumptions and the problem no longer bears even a passing resemblance to the real world; too few and there may be no apparent way to get a solution. 

We can make progress by starting with a small piece of the problem:  What’s the probability that any men’s college basketball game would end with a score of 76-68?  A quick look at widely available results of games reveals that teams typically score between 50 to 90 points.  So a very simple first guess is to assume that scores are uniformly distributed in that range, thus a score of 76-68 would have a probability of about $(1/40)^2 = 1/1600$.  (Note that we have just snuck in another assumption that the two scores are independent!)

The probability that any score would happen three times in a row, assuming independence between games as well as between teams, would be about $(1/1600)^3$, or about once in every four billion three-game sequences.  Wow, that IS a rare event!  But wait a minute\ldots the interesting thing here is that the same score happened three times in a row; thus the score of the first game doesn’t really matter!  The problem is just as interesting if all of the games had been 90-58 or 63-62.  Thus a better estimate might by $(1/1600)^2$, about once in every 2.5 million three-game sequences.  With about 2000 college teams playing about 30 games per season, this is apparently about a ``once or twice in a lifetime'' event.  To know if this quick guess is reasonable, it should be tested against real data, but with over 2000 men's and women's games played at many levels, the necessary data collection would be a challenge.

When Prof.~Staab heard Prof.~Cleary describe the same score problem he realized that we could, in a slightly different setting, work the other way around by moving to a game with lots of accessible historical data, Major League Baseball. We first examine the historical record of the modern-era in Major League Baseball (1901-2019) and use the data to answer the question of the likelihood of a same-score streak of length 3. Then, using a probability distribution that has been proposed for MLB scores, we run a simulation to see if the model produces score streaks at the rate we'd expect.

%\section{More Stuff}
%
%One of the difficulties in answering this question is the nature of basketball in the United States.  There are a lot of games played a many levels and collecting all of the basketball scores to test historically if this has happened is a challenge in data collection.  But this is an interesting question, and we investigate it first with a game with lots of historical date, Major League Baseball.  We first examine the historical record of the modern-era in Major League Baseball (1900-2019) and answer the question of the likelihood of a same-score streak of 3.  We then perform a simulation of major league baseball for same-score streaks to help understand the underlying mathematics of a score streak.  
%
%There are plenty of streak studies in sports, however, they consist of references to lengths distributions of streaks \cite{albright:2012}. 

% Here's another reference\cite{ribeiro:2016}

\section{What is a score streak?}

With the driving example of the Barton College streak, we need a careful definition of what it means to be a score streak.  We use the following:

\begin{definition*}
A \textbf{same-score streak of order $n$} or an \textbf{order-$n$ streak} is a sequence of $n$ consecutive games in a team's season such that the scores of the all games in the sequence are the same.  The team should have the same score in each game and the opponents' score should be the same in each game. 
\end{definition*}

The Barton College example is an order-3 streak. However, the sequence of three games between teams A and B with the scores: A--2, B--3;  A--2, B--3; A--3, B--2,  is not an order-3 streak due to the third game in which the score is identical, but the teams' scores are reversed.  The first two games, however, form an order-2 streak for each team.  

Here's an example from the 2019 Major League Baseball (MLB) season: 
\begin{center}
\begin{tabular}{lllll} 
Date & Team A & Score & Team B & Score \\ \hline
June 10, 2019 & L.A. Angels & 5 & L.A. Dodgers & 3 \\
June 11, 2019 & L.A. Angels & 5 & L.A. Dodgers & 3 \\
June 13, 2019 & L.A. Angels & 5 & Tampa Bay Rays & 3
\end{tabular}
\end{center}

This is an example of an order-3 streak for the Angels (as well as two order-2 streaks).  Additionally, there is an order-2 streak for the L.A. Dodgers.  

An exercise for the reader:  Imitating our calculation for college basketball, what is a rough estimate for how often consecutive MLB games would have the same score?

\section{Historical Streaks in Major League Baseball}

With 119 years of data available for what is deemed the modern era of Major League Baseball (MLB), we sift through all games of these years and apply the definition to see what comes out. The website \href{http://baseball-reference.com}{baseball-reference.com} has in-depth level of information about nearly every major league baseball game ever played.  For this study, we downloaded the results of every regular-season game between the 1901 and 2019 seasons, for a total of 199,692 games.  For each game, the date, the home and away teams as well as each score was recorded.  
 
A same-score streak of order 2 is quite common in baseball, due to both the large number of games in a given season and the relatively low variance of scores in MLB games.  For example, in 2019, there were 56 order-2 streaks (and three of these were involved in the example order-3 shown above). 

The bar chart below shows the number of order-2 streaks in a given year from 1901 to 2019.  From this figure, you can see that the total number of same-score games in a given year range from just under 20 to over 80 and there is an upward trend over the years.  The increase in number of streaks per season can be explained by two factors: the increase in the number of games played per season and the increase in the number of teams over the years.  In 1901-1903, only 140 games were played each year, from 1904--1960, 154 games and since 1961, there have been 162 games played. The plot demonstrates these trends with color-coding the bar with the number of teams included.  

\begin{center}
\pgfplotsset{scale=0.55}
\input{plots/order2_per_year_num_teams.tex}
\end{center}

%The main reason for the smaller number of same-score streaks for the first 60 years of the historical data is that the number of teams playing, the season-length and therefore the total number of games has not be constant.  The following table lists the number of teams, games per season and total number of games played for each season between 1901 and 2017.  
%
%\begin{center}
%\begin{tabular}{rccc} 
%years & number of teams & season length & max number of games played per year \\ \hline
%1901-1903 & 16 & 140 & 1120\\
%1904-1913 & 16 & 154 & 1232\\
%1914-1915 & 24 & 154 & 1848 \\
%1916-1960 & 16 & 154 & 1232 \\
%1961 & 18 & 162 & 1458  \\
%1962-1968 & 20 & 162 & 1620 \\
%1969-1976 & 24 & 162 & 1944\\
%1977-1992 & 26 & 162 & 2106\\
%1993-1997 & 28 & 162 & 2268 \\
%1998-2019 & 30 & 162 & 2430\\
%\end{tabular}
%\end{center}

Note: often, the actual number of games played per year is lower than the max due to rain out games that were not rescheduled.  Two years, 1981 and 1994, had seasons shortened substantially due to labor issues; these seasons are visible outliers in some of our graphs.

Hopefully you, the reader, took some time above to estimate the probability of an order-2 streak.  The historic number of order-2 streaks in MLB is 5,313 and the total number of pairs of games for each team is 141,397.  This results in a probability of 0.0376 that any pair of games is an order-2 streak.  So how did you do with your estimation? 

The following two plots are alternative ways of examining the scores of all games in order-2 streaks: 

\begin{center}
\includegraphics[width=7in]{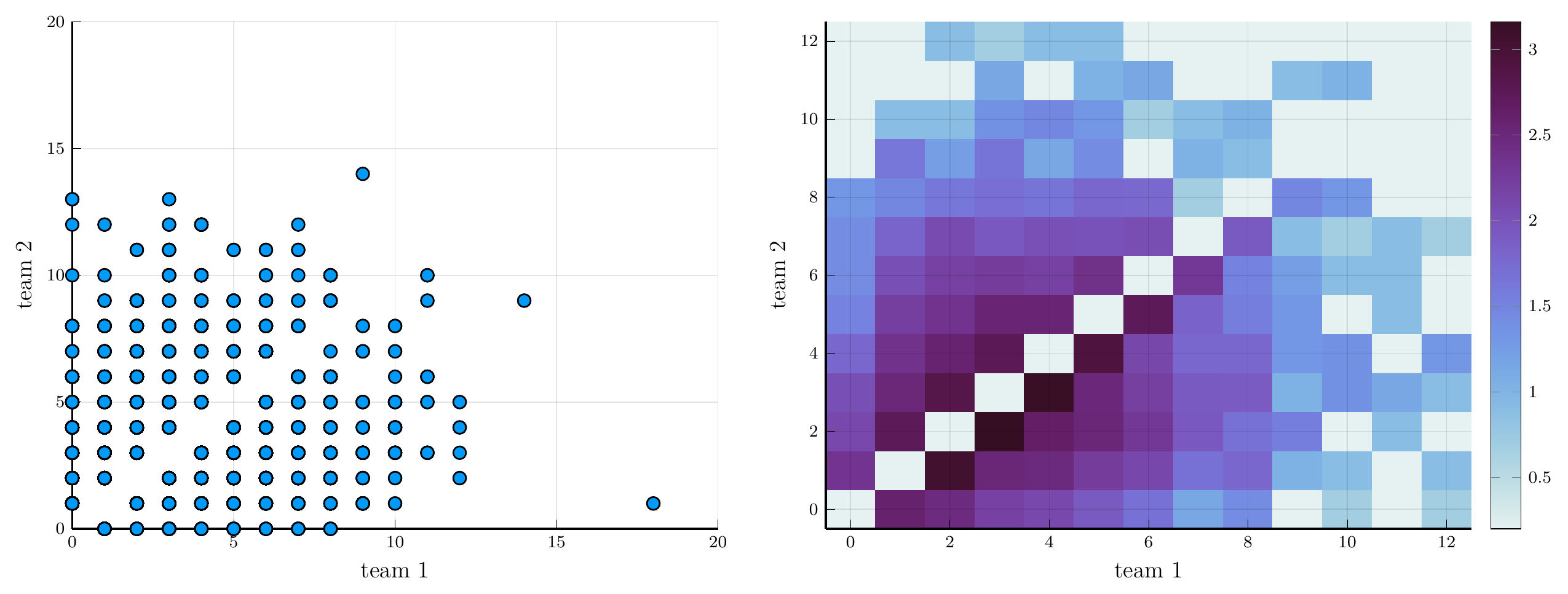}
\end{center}
The plot on the left is a scatter plot of all order-2 streak games, which each dot represents one or more games with that score.  The plot on the right is a heat map which gives the relative number of games with that score.  The color is based on a base-10 log scale and for reference there are 459 games where team 1 beats team 2 with a score of 3-2. 

In both plots, the main diagonal is empty since baseball games do not end in a tie.  The point in the scatter plot in the lower right is where the New York Giants beat the Philadelphia Phillies 18-1 in back-to-back games on August 2 and 4 of 1933.  

\subsection{Streaks of Order 3 and 4}

Not surprisingly, order-3 streaks are much less probable than the corresponding order-2 streaks. There were 134 such streaks over the 119-year timespan that is studied.  The following bar plot shows the number of games per year. 
\begin{center}
\input{plots/order3_per_year.tex}
\end{center}

The chart shows between 0 and 6 order-3 streaks per year, with the 2015 season having a seemingly outlier-year of 6. However, unlike the order-2 streaks, there is no clear upward trend over the years. 

\subsection{Same Score Streaks are Lower Scoring than Average Games}

If we compare the total number of runs in an order-2 or order-3 streaks to that of all games, we can see that they are lower scoring.  The following bar graph shows this where the 3 distributions are compared.

\begin{center}
\input{plots/total_run_distribution.tex}
\end{center}

This plot also shows that an even number of total runs is less common than odd totals of similar magnitude, the discrepancy accounted for by ``walk-off'' games when the home team wins by one-run.

And the following table shows a comparison of the mean and standard deviation for the total number of runs in a single game in a same-score streak. 

\begin{center}
\begin{tabular}{r|c|c|c} 
& all games & order-2 & order-3  \\ \hline
mean & 8.81 & 6.84 & 5.79 \\
std. dev. & 4.60 & 3.29 & 2.70 
\end{tabular}
\end{center}

The results show that both the mean and standard deviation shifts lower for higher order streaks relative to all games.  

Finally, we note that there have been just three streaks of order 4 in MLB history.  The Washington Senators lost 0-2 in 4 consecutive games in 1958.  The San Francisco Giants lost 2-3 in 4 consecutive games in 1961.  More recently, from June 4 to June 7 in 2008, San Diego Padres won 4 consecutive games with the score of 2 to 1.  The first game was against the Chicago Cubs and the latter 3 were against the New York Mets. This 4-game streak accounted for 6\% of the Padres' wins in the 2008 season. Also, due to the way streaks are counted, there were three order-3 streaks (two for the Padres and one for the Mets) and five order-2 streaks (three for the Padres and two for the Mets).  All three of the order-3 streaks in the 2008 season were associated with this order-4 streak.

\section{Fitting the Data to a Weibull Distribution} \label{sect:weibull}

 Now that we have the empirical data, let's see if simulated baseball seasons using a proposed distribution for MLB runs match up well with our observations. Miller \cite{miller:2007} used a Weibull distribution to model run distributions of American League teams in 2004 to provide a theoretical justification for Bill James' Pythagorean Formula \cite{james:1983}.  As was found, the Weibull distribution can be fit well to such distributions and we use it to model home and away run distribution for all MLB seasons from 1901 to 2019.  

A Weibull distribution \cite{peacock:2013}, is a three-parameter function of the form:
\begin{align*}
f(x; \alpha, \beta, \gamma) & = \begin{cases}
\frac{\gamma}{\alpha} \left(\frac{x-\beta}{\alpha}\right)^{\gamma-1} e^{-((x-\beta)/\alpha)^{\gamma}}, & x \geq \gamma, \\
0, & x < \gamma
\end{cases}
\end{align*}
and is clearly a continuous distribution. Since the number of runs scored in a game is discrete, we discretize the distribution as follows:
\begin{align*}
P(r) = P(X = r \text{~runs}) = F(r+1)-F(r) & = \int_r^{r+1} f(x;\alpha,\beta,\gamma) \, dx 
\end{align*}
where $F$ is the c.d.f of $f$ or simply the antiderivative and finding it is a fun exercise in integration by substitution.  Give it a try. 

Let $R_i$ be the historic percentage of games where the away team scores $i$ runs.  We define
\begin{align}
S_a(\alpha_a, \beta_a,\gamma_a) & = \sum_{i=0}^{30} (P(i) - R_i)^2
\label{eq:S}
\end{align}
where the upper bound of 30 is chosen to be the maximum number of runs a team has scored in a single game historically and we also define a corresponding function $S_h$ similar to (\ref{eq:S}) for home games as well.   We then use the JuMP package of the Julia language\footnote{JuMP (\texttt{http://www.juliaopt.org/JuMP.jl/latest/}) is modeling language useful for optimization problems and Julia (\texttt{http://julialang.org}) is a general programming language designed for scientific computing.} to minimize $S_a$ and $S_h$ and determine that
\begin{align*}
% k_a & = 1.5817, & \lambda_a & =5.390, & k_h &= 1.715, & \lambda_h & = 5.557 \\
\alpha_a &= 1.674, &\beta_a & =   5.564,&  \gamma_a & =  -0.231 &  
\alpha_h &= 1.832, &\beta_h & =   5.787,&  \gamma_h & =  -0.287 
\end{align*}
where the $a$ and $h$ subscripts are the parameters for the away and home teams respectively.  A visual check on the accuracy of the fit can be seen with the following plot of the historic and model of the away run distribution. 
\begin{center}
\input{plots/weibull-historic.tex}
\end{center}

\section{Simulating Same-Score Streaks}

We look to simulate same-score streaks by producing games with randomized scores and then checking the number of streaks.  There are two important aspects of the simulation which arise directly from the definition of a same-score streak. 
\begin{enumerate}
\item When a random game is generated, what distribution should be used?
\item When determining a streak, it is important that the games have a certain order to them.  In short this is a schedule and the way it is constructed is important. 
\end{enumerate}

\subsection{Game Score Distribution} \label{sect:dist}

We developed two models based on the Weibull distribution. The \textbf{simple Weibull distribution} selects the home and away scores randomly based on the distribution developed in section \ref{sect:weibull}, where any tie score is thrown out and a new draw is made.  

Although is relatively straightforward approach of picking from the two distributions assumes independence and in baseball (and most sports), run distribution cannot be independent because of the fact that there are no ties in baseball.  From a statistics point of view, once the number of runs is ``chosen'' from any random distribution for one team, the second teams scored is restricted, so it cannot be independent. 

Instead, we develop a bivariate distribution. The following plot shows a portion of the historic run distribution that we wish to model
\begin{center}
\includegraphics[width=3in]{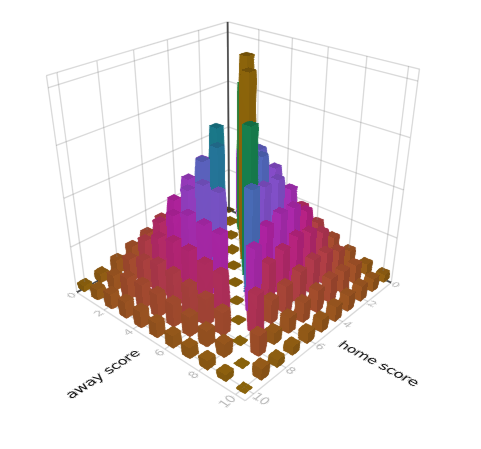}
\end{center}
The ``canyon'' that is see in the middle is where the two teams tie (which again, happens very rarely). Also, we see that on the home side of the canyon the fraction of games is the highest.  This is due to the fact that one-run games are relatively prevalent and in these the home team wins often. 

To develop a model of this distribution, we fit a Weibull to the diagonals where $r_h-r_a=k$, for $k$ some constant in a manner similar to that in section \ref{sect:weibull}.  Thus, this model is 58 Weibull distributions, weighted to match the actual fraction of games with the run difference.  Although a bit complicated, the take-home message is that we can create a bivariate distribution to determine the scores of baseball games without ties.  We will call this the \textbf{bivariate Weibull distribution}.

\subsection{Modeling a Schedule} \label{sect:simple:schedule}
This section explains how one can model a generic sports schedule, but is applicable to MLB.  Assume that there are $T$ teams in a league and each team plays $G$ games in a season. Overall, there are a total of $GT/2$ games.  A \textbf{schedule} will be a sequence of pairs of teams, such that for each pair $(i,j)$ the home team is the $i$th team and the away team is the $j$th team.  For each game, $i\neq j$ and in the schedule each team will have $G/2$ home and $G/2$ away games.  If the schedule is produced randomly with just these properties, we will call this a \textbf{basic schedule}.

%The results of the basic schedule are poor, therefore we didn't include them in this paper.  An improvement includes both home and away teams for each game.  The schedule is created to ensure that each team as an equal number of home and away games.  Such a schedule is called a \textbf{better schedule}.

However this schedule results in fewer score streaks of order 2 and 3 than the historical evidence shows.  The reason for this could be that series play a huge role in baseball and increases the number of score streaks.  Looking at the order-3 streak in Section 2, this included two other order-2 streaks.  The order-2 streak for the L.A. Dodgers is due to the fact that there was a series played.  If the L.A. Angels played a different team in each game, these extra streaks wouldn't arise.  

In addition to the properties above, we ensure that each team plays another team in either a 2-, 3- or 4-game series distribution in a manner similar to the MLB schedule. We will call this a \textbf{realistic schedule}\footnote{The details of how the actual MLB schedule is created is fascinating.  It has been for quite some time created by a husband-wife team using a combination of computer code and some fancy by-hand juggling techniques. An ESPN film for the series 30 for 30 featured them. \cite{schedule:2015}}.

% A schedule where every team has $G/2$ home and $G/2$ away games and plays 4-, 3-, 2-game and occasional 1-game series will be called a   The main difference between schedules produced for this study and actual MLB schedule is we don't care about dates, number of teams within divisions and geography, whereas actual schedules are developed with these in mind.  

\section{Simulation Same Streaks using a Model}

To simulate a season (or 119 seasons), we develop a schedule, simulate a score for each game and count the streaks. Although there a numerous ways to do this, we will show two such models:

\subsection{Simple Weibull Model with Basic Schedule}

For this model, the \emph{simple Weibull distribution} is used to generate game scores and the \emph{basic schedule} is used for counting streaks. For each season, a schedule is developed for the historic number of teams and games, then each game is modeled randomly.  We repeat this process for 10,000 seasons per year and the plot below shows the minimum, mean, maximum as well as 5th and 95th percentile of order-2 streaks. 

\begin{center}
\input{plots/sim2-comparison.tex}
\end{center}
The actual number of historic order-2 streaks often exceeds that of this model's simulated trials maximum, which indicates that most likely this is not capturing the actual situation.   For some quantitative comparison, there are 10 years where the historic number of streaks in a season exceeds the max of the simulation and there are 89 years (or 75\%) which exceed the simulation mean.  In short, this simulation is not capturing the historic streaks.  

% \subsection{Bivariate Weibull with a Better Schedule}

% As stated above, the Univariate Weibull distribution cannot accurately model the runs scored in a game. For this simulation we used the bivariate Weibull distribution described above in Section \ref{sect:dist}.  The results are shown in the plot: 
% \begin{center}
% %\includegraphics[width=6in]{../paper/plots/sim3-comparison.pdf}
% \input{plots/sim3-comparison.tex}
% \end{center}
% which appear better again, however, note that in 70 years (59\%) the historic number of streaks exceeds the simulation mean and in 29 years, the number of streaks exceeds the 95th percentile, where this should occur on average in 6 years.  

\subsection{Bivariate Weibull with a Realistic Schedule}

In this model, the \emph{bivariate Weibull distribution} is used along with the \emph{realisitic schedule} for each of the 119 seasons and this is simulated 10,000 times per season.  The results are shown in the plot: 
\begin{center}
\input{plots/sim4-comparison.tex}
\end{center}

In 67 years (56\%), the historic number of streaks exceeds the simulation mean and in only  20 years (17\%), the number of streaks exceeds the 95th percentile.  There are 5 years (or 4\%) with the fewer than the 5th percentile of the simulation number of streaks.  In no years does the historic number exceed the simulation maximum or is below the simulation minimum.  Although simulation distribution is not quite that of the historic data, this model seems to perform the best of those studied.  

\subsection{Simulated Order-3 and Order-4 streaks} 

We used the the \emph{bivariate Weibull distribution} and the \emph{realisitic schedule} to model order-3 and order-4 streaks.  As with previous models $10,000$ simulations were run for each year for the number of teams and games consistent with that year.  The following plot is generated:
\begin{center}
\input{plots/order3-sim-comparison.tex}
\end{center}

In nearly half of the years (64/119), the historic number of order-3 streaks exceeds the simulation mean.  And in 7.5\% (9 of 119) of the years, the historic number of order-3 streaks exceeds the 95th percentile--although if you include when they equal that it is 22\% (26 of 119).   Again, since the historic data fits well within the simulation, this shows that the model reasonable captures the order-3 streak.

We found that there have been three order-4 streaks in the history of Major League Baseball.  We also performed simulations to determine how likely this is to occur.  On any given year, an order-4 streak is quite unlikely, instead, we simulated 119 years using the model from the order-3 streaks.    We then summed the results over 119 years.  This was done 10,000 times and the resulting plot shows the number of order-4 streaks throughout a simulated 119-year set of seasons:

\begin{center}
\begin{tikzpicture}[/tikz/background rectangle/.style={fill={rgb,1:red,1.0;green,1.0;blue,1.0}, draw opacity={1.0}}, show background rectangle]
\begin{axis}[title={}, title style={at={{(0.5,1)}}, font={{\fontsize{14 pt}{18.2 pt}\selectfont}}, color={rgb,1:red,0.0;green,0.0;blue,0.0}, draw opacity={1.0}, rotate={0.0}}, legend style={color={rgb,1:red,0.0;green,0.0;blue,0.0}, draw opacity={1.0}, line width={1}, solid, fill={rgb,1:red,1.0;green,1.0;blue,1.0}, fill opacity={1.0}, text opacity={1.0}, font={{\fontsize{8 pt}{10.4 pt}\selectfont}}, at={(1.02, 1)}, anchor={north west}}, axis background/.style={fill={rgb,1:red,1.0;green,1.0;blue,1.0}, opacity={1.0}}, anchor={north west}, xshift={1.0mm}, yshift={-1.0mm}, width={145.4mm}, height={74.2mm}, scaled x ticks={false}, xlabel={number of simulated order-4 streaks/119-year season}, x tick style={color={rgb,1:red,0.0;green,0.0;blue,0.0}, opacity={1.0}}, x tick label style={color={rgb,1:red,0.0;green,0.0;blue,0.0}, opacity={1.0}, rotate={0}}, xlabel style={, font={{\fontsize{11 pt}{14.3 pt}\selectfont}}, color={rgb,1:red,0.0;green,0.0;blue,0.0}, draw opacity={1.0}, rotate={0.0}}, xmajorgrids={true}, xmin={-1.06744}, xmax={11.06744}, xtick={{0.0,2.0,4.0,6.0,8.0,10.0}}, xticklabels={{$0$,$2$,$4$,$6$,$8$,$10$}}, xtick align={inside}, xticklabel style={font={{\fontsize{8 pt}{10.4 pt}\selectfont}}, color={rgb,1:red,0.0;green,0.0;blue,0.0}, draw opacity={1.0}, rotate={0.0}}, x grid style={color={rgb,1:red,0.0;green,0.0;blue,0.0}, draw opacity={0.1}, line width={0.5}, solid}, axis x line*={left}, x axis line style={color={rgb,1:red,0.0;green,0.0;blue,0.0}, draw opacity={1.0}, line width={1}, solid}, scaled y ticks={false}, ylabel={fraction}, y tick style={color={rgb,1:red,0.0;green,0.0;blue,0.0}, opacity={1.0}}, y tick label style={color={rgb,1:red,0.0;green,0.0;blue,0.0}, opacity={1.0}, rotate={0}}, ylabel style={, font={{\fontsize{11 pt}{14.3 pt}\selectfont}}, color={rgb,1:red,0.0;green,0.0;blue,0.0}, draw opacity={1.0}, rotate={0.0}}, ymajorgrids={true}, ymin={-0.007431999999999999}, ymax={0.25516533333333336}, ytick={{0.0,0.05,0.1,0.15000000000000002,0.2,0.25}}, yticklabels={{$0.00$,$0.05$,$0.10$,$0.15$,$0.20$,$0.25$}}, ytick align={inside}, yticklabel style={font={{\fontsize{8 pt}{10.4 pt}\selectfont}}, color={rgb,1:red,0.0;green,0.0;blue,0.0}, draw opacity={1.0}, rotate={0.0}}, y grid style={color={rgb,1:red,0.0;green,0.0;blue,0.0}, draw opacity={0.1}, line width={0.5}, solid}, axis y line*={left}, y axis line style={color={rgb,1:red,0.0;green,0.0;blue,0.0}, draw opacity={1.0}, line width={1}, solid}, colorbar style={title={}}, point meta max={nan}, point meta min={nan}]
    \addplot[color={rgb,1:red,0.0;green,0.0;blue,0.0}, name path={48b611ab-6bf4-449c-a786-65c7b90f8ce8}, draw opacity={1.0}, line width={1}, solid, fill={rgb,1:red,0.0;green,0.6056;blue,0.9787}, fill opacity={1.0}, area legend, forget plot]
        coordinates {
            (-0.4,0.1442)
            (-0.4,0.0)
            (0.4,0.0)
            (0.4,0.1442)
            (-0.4,0.1442)
        }
        ;
    \addplot[color={rgb,1:red,0.0;green,0.0;blue,0.0}, name path={48b611ab-6bf4-449c-a786-65c7b90f8ce8}, draw opacity={1.0}, line width={1}, solid, fill={rgb,1:red,0.0;green,0.6056;blue,0.9787}, fill opacity={1.0}, area legend, forget plot]
        coordinates {
            (0.6,0.24713333333333334)
            (0.6,0.0)
            (1.4,0.0)
            (1.4,0.24713333333333334)
            (0.6,0.24713333333333334)
        }
        ;
    \addplot[color={rgb,1:red,0.0;green,0.0;blue,0.0}, name path={48b611ab-6bf4-449c-a786-65c7b90f8ce8}, draw opacity={1.0}, line width={1}, solid, fill={rgb,1:red,0.0;green,0.6056;blue,0.9787}, fill opacity={1.0}, area legend, forget plot]
        coordinates {
            (1.6,0.24773333333333333)
            (1.6,0.0)
            (2.4,0.0)
            (2.4,0.24773333333333333)
            (1.6,0.24773333333333333)
        }
        ;
    \addplot[color={rgb,1:red,0.0;green,0.0;blue,0.0}, name path={48b611ab-6bf4-449c-a786-65c7b90f8ce8}, draw opacity={1.0}, line width={1}, solid, fill={rgb,1:red,0.0;green,0.6056;blue,0.9787}, fill opacity={1.0}, area legend, forget plot]
        coordinates {
            (2.6,0.1722)
            (2.6,0.0)
            (3.4,0.0)
            (3.4,0.1722)
            (2.6,0.1722)
        }
        ;
    \addplot[color={rgb,1:red,0.0;green,0.0;blue,0.0}, name path={48b611ab-6bf4-449c-a786-65c7b90f8ce8}, draw opacity={1.0}, line width={1}, solid, fill={rgb,1:red,0.0;green,0.6056;blue,0.9787}, fill opacity={1.0}, area legend, forget plot]
        coordinates {
            (3.6,0.09906666666666666)
            (3.6,0.0)
            (4.4,0.0)
            (4.4,0.09906666666666666)
            (3.6,0.09906666666666666)
        }
        ;
    \addplot[color={rgb,1:red,0.0;green,0.0;blue,0.0}, name path={48b611ab-6bf4-449c-a786-65c7b90f8ce8}, draw opacity={1.0}, line width={1}, solid, fill={rgb,1:red,0.0;green,0.6056;blue,0.9787}, fill opacity={1.0}, area legend, forget plot]
        coordinates {
            (4.6,0.05026666666666667)
            (4.6,0.0)
            (5.4,0.0)
            (5.4,0.05026666666666667)
            (4.6,0.05026666666666667)
        }
        ;
    \addplot[color={rgb,1:red,0.0;green,0.0;blue,0.0}, name path={48b611ab-6bf4-449c-a786-65c7b90f8ce8}, draw opacity={1.0}, line width={1}, solid, fill={rgb,1:red,0.0;green,0.6056;blue,0.9787}, fill opacity={1.0}, area legend, forget plot]
        coordinates {
            (5.6,0.0236)
            (5.6,0.0)
            (6.4,0.0)
            (6.4,0.0236)
            (5.6,0.0236)
        }
        ;
    \addplot[color={rgb,1:red,0.0;green,0.0;blue,0.0}, name path={48b611ab-6bf4-449c-a786-65c7b90f8ce8}, draw opacity={1.0}, line width={1}, solid, fill={rgb,1:red,0.0;green,0.6056;blue,0.9787}, fill opacity={1.0}, area legend, forget plot]
        coordinates {
            (6.6,0.0106)
            (6.6,0.0)
            (7.4,0.0)
            (7.4,0.0106)
            (6.6,0.0106)
        }
        ;
    \addplot[color={rgb,1:red,0.0;green,0.0;blue,0.0}, name path={48b611ab-6bf4-449c-a786-65c7b90f8ce8}, draw opacity={1.0}, line width={1}, solid, fill={rgb,1:red,0.0;green,0.6056;blue,0.9787}, fill opacity={1.0}, area legend, forget plot]
        coordinates {
            (7.6,0.003)
            (7.6,0.0)
            (8.4,0.0)
            (8.4,0.003)
            (7.6,0.003)
        }
        ;
    \addplot[color={rgb,1:red,0.0;green,0.0;blue,0.0}, name path={48b611ab-6bf4-449c-a786-65c7b90f8ce8}, draw opacity={1.0}, line width={1}, solid, fill={rgb,1:red,0.0;green,0.6056;blue,0.9787}, fill opacity={1.0}, area legend, forget plot]
        coordinates {
            (8.6,0.0013333333333333333)
            (8.6,0.0)
            (9.4,0.0)
            (9.4,0.0013333333333333333)
            (8.6,0.0013333333333333333)
        }
        ;
    \addplot[color={rgb,1:red,0.0;green,0.0;blue,0.0}, name path={48b611ab-6bf4-449c-a786-65c7b90f8ce8}, draw opacity={1.0}, line width={1}, solid, fill={rgb,1:red,0.0;green,0.6056;blue,0.9787}, fill opacity={1.0}, area legend, forget plot]
        coordinates {
            (9.6,0.00046666666666666666)
            (9.6,0.0)
            (10.4,0.0)
            (10.4,0.00046666666666666666)
            (9.6,0.00046666666666666666)
        }
        ;
\end{axis}
\end{tikzpicture}
\end{center}

The historic number of three order-4 streaks is quite likely (17\%) from this simulation, although in any given year having one is quite low.  From these simulations, the probability that an order-4 streak occurred in any given year was only 1.7\%,  having an order-4 streak in a 119-year timespan is not rare.  
 
\section{Conclusion}

We looked at the past 119 years of Major League Baseball to examine the number of same-score streaks of length 2, 3 and 4.  In trying to explain streak frequency, we modeled scores of baseball games using both a simple Weibull distribution as well as a bivariate Weibull distribution that both quantitatively and visually appear to fit the historic data well.  In addition, we studied two different scheduling models and noted that using more realistic models was important in capturing the correct number of same-score streaks.  Of all the models, the realistic schedule which includes series of games similar to that of actual MLB schedules seemed to perform best when used with the bivariate Weibull distribution.  

In a larger sense, we hope we have given a demonstration of how probability modeling works in practice.  A problem of interest comes to our attention.  We think about how to model the event in a simple way, and if possible we collect related data.  Then we iterate between data and model until we have a result that fits the data well and sheds some light on the underlying process. 

\bibliographystyle{plain}
\bibliography{ref}

\end{document}